\documentclass[]{spie}  

\usepackage{amsmath,amsfonts,amssymb}
\usepackage{graphicx}

\title{Maximizing the Information Gain of a Single Ion Microscope using Bayes Experimental Design}

\author[a]{Georg Jacob}
\author[a]{Karin Groot-Berning}
\author[a]{Ulrich G. Poschinger}
\author[a]{Ferdinand Schmidt-Kaler}
\author[b]{Kilian Singer}
\affil[a]{QUANTUM, Institut f\"ur Physik, Universit\"at Mainz, Staudingerweg 7, 55128 Mainz, Germany}
\affil[b]{Experimentalphysik I, Institut f\"ur Physik, Universit\"at Kassel, Heinrich-Plett-Stra{\ss}e 40, 34132 Kassel, Germany}

\authorinfo{Further author information: (Send correspondence to G.J.)
\\G.J.: E-mail: georg.jacob@uni-mainz.de, Telephone: +49 (0)6131 39 23671
\\  K.S.: E-mail: ks@uni-kassel.de, Telephone: +49 (0)561 804-4235
\\  F.S.K.: E-mail: fsk@uni-mainz.de, Telephone: +49 (0)6131 39 26234}

\begin{document} 
\maketitle

This paper has been published in Proc. SPIE 9900, Quantum Optics, 99001A (April 29, 2016); Copyright 2009 Society of Photo Optical Instrumentation Engineers. One print or electronic copy may be made for personal use only. Systematic electronic or print reproduction and distribution, duplication of any material in this paper for a fee or for commercial purposes, or modification of the content of the paper are prohibited.

\begin{abstract}

We show nanoscopic transmission microscopy, using a deterministic single particle source and compare the resulting images in terms of signal-to-noise ratio, with those of conventional Poissonian sources. Our source is realized by deterministic extraction of laser-cooled calcium ions from a Paul trap. Gating by the extraction event allows for the suppression of detector dark counts by six orders of magnitude. Using the Bayes experimental design method, the deterministic characteristics of this source are harnessed to maximize information gain, when imaging structures with a parametrizable transmission function. We demonstrate such optimized imaging by determining parameter values of one and two dimensional transmissive structures.
\end{abstract}

\keywords{Ion trapping, Laser cooling, Charged-particle beams, Transmission microscopy, Ion implantation, Information gain optimized imaging.}

\section{INTRODUCTION}
\label{sec:intro}  

Trapping and laser-cooling of single ions, has enabled substantial advance in applications as diverse as quantum information processing~\cite{blatt2008entangled}, creation of precise clocks~\cite{rosenband2008frequency,ludlow2015optical} and fundamental research on quantum phenomena\cite{Wineland2013NobelLecture}. Here we present a new application based on laser-cooled ions which extends the boundaries of nanoscopic imaging to the single particle limit. We implement a transmission microscope using an intrinsically deterministic ion source based on extracting single laser-cooled $^{40}$Ca$^+$ ions from a linear segmented Paul trap~\cite{schnitzler2009deterministic,izawa2010controlled}. This offers several advantages over microscopy with conventional particle sources - in particular, a higher signal-to-noise ratio~(SNR), especially at low exposures and an higher information gain per particle. This can be increased even further, when combined with the Bayes experimental design method, which allows for maximizing the information gain for each individual particle probe.

Our apparatus was conceived for ion implantation as well, because besides the favourable statistical properties, it also provides an ultracold monochromatic ion beam where ultimately the phase space occupation in transversal and longitudinal direction is limited by the Heisenberg uncertainty principle. Exactly the same characteristics renders it also favourable for imaging. These different applications are highly complementary, since microscopy - or more precisely alignment by imaging of the sample - is essential for an accurate absolute positioning of dopants, free of parallax errors.

\section{Single ion microscope setup}

\begin{figure}[ht]
\begin{center}
\hfill
\includegraphics[height=4cm]{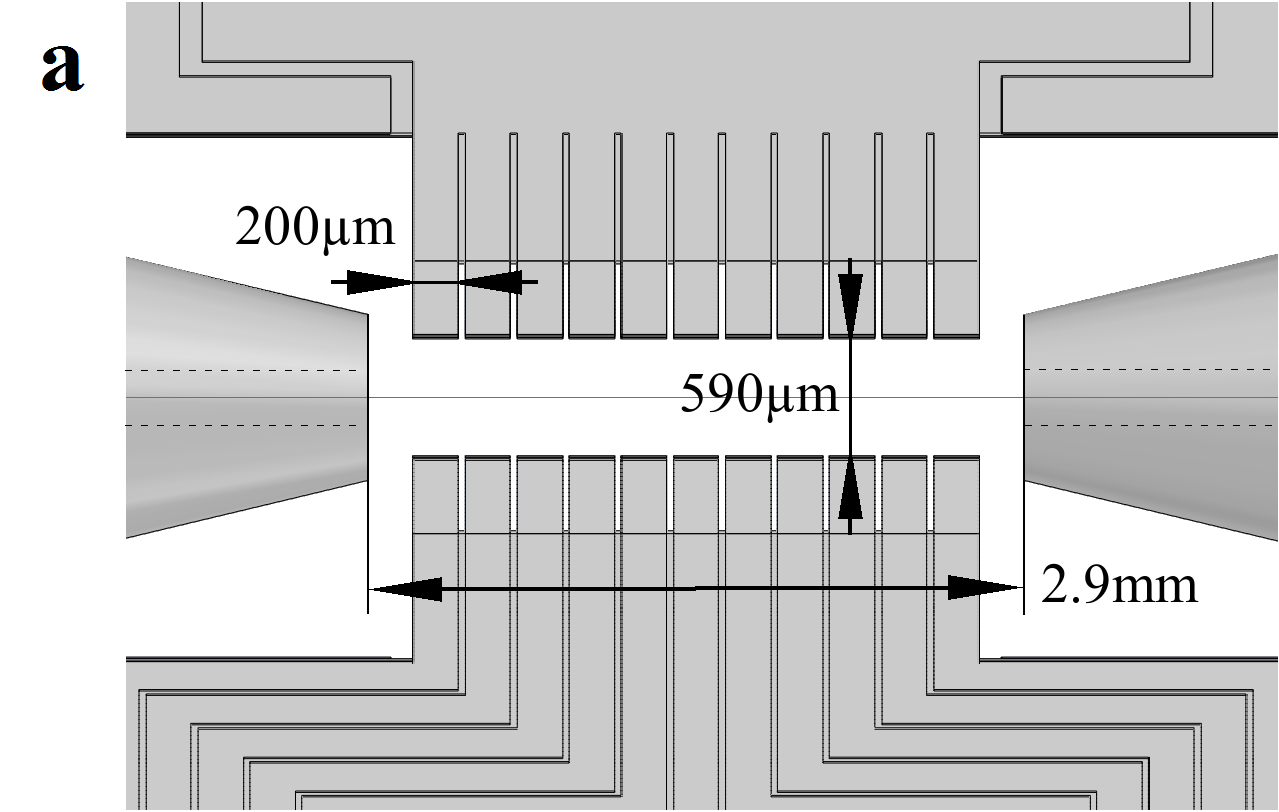}
\hfill
\includegraphics[height=3.9cm]{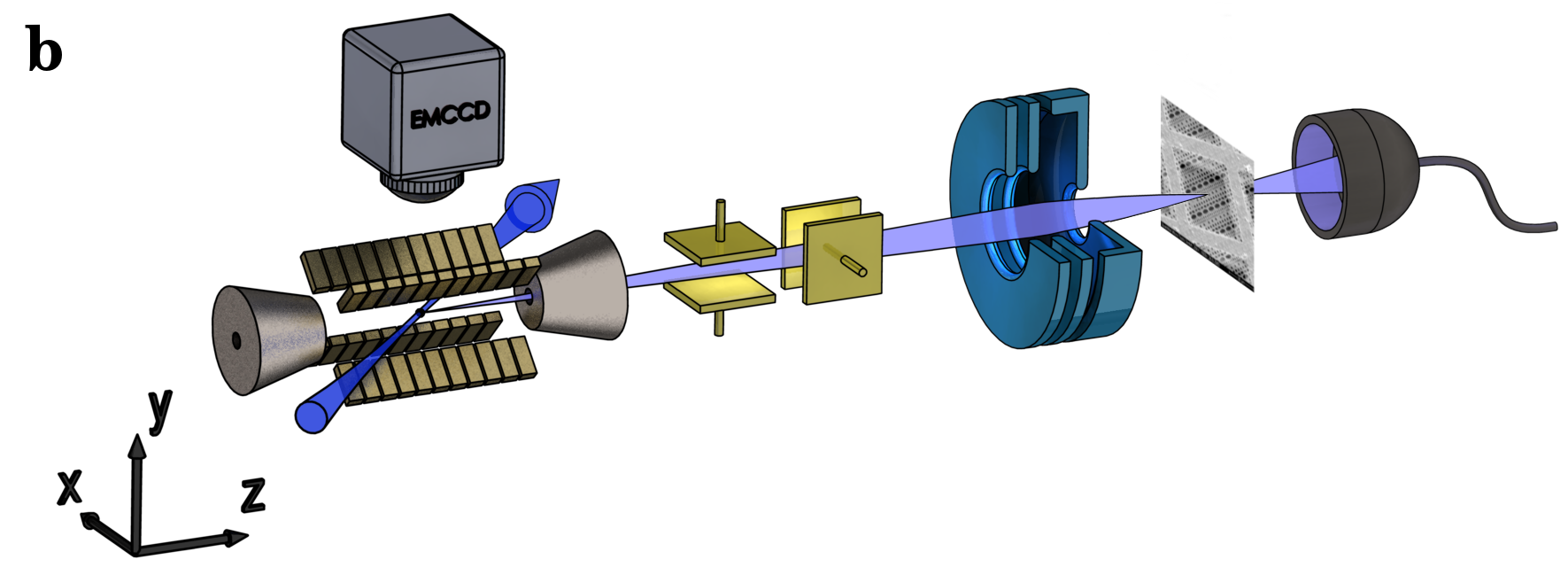}
\hfill
\caption{a) True to scale image of the trap geometry denoting important distances, as seen from the EMCCD-camera. b) Sketch of the single-ion implanter and microscope. The ion trap consists of segmented electrodes and end-caps. Laser-cooling (blue arrow) and imaging system (EMCCD) are shown. Deflection electrodes (yellow) and einzel-lens (blue) are placed along the ion extraction pathway (light-blue). In the focal plane a profiling edge or alternatively a transmissive mask is placed on a three-axis nano-translation stage (not shown). Single ion counting is performed with a secondary electron multiplier device.\label{fig:trap}}
\end{center}
\end{figure}

The experimental setup is based on a Paul trap, comprising four micro-fabricated alumina chips, which are arranged in an X-shaped configuration and two pierced end-caps (see Figure~\ref{fig:trap}a). Each chip consists of 11 electrodes to shape the potential in the axial direction. The trap is operated at frequencies $\omega/(2\pi)=0.58\,$MHz and 1.4$\,$MHz for the axial and radial directions, respectively. Calcium ions are generated by photo-ionization and laser cooled on the S$_{1/2}$ to P$_{1/2}$ dipole transition. Other ion species are created by a commercial ion gun~\footnote{Specs, Ion Source IQE 12/38} which is attached to the vacuum chamber. This allows for direct loading of ions into the trap along the axial direction.

In order to achieve a high repetition rate of the ion extraction, an automated loading of a predefined number of ions is implemented. Initially a random number of ions is trapped and Doppler-cooled using laser light near 397$\,$nm. The ion number is counted by imaging the ion fluorescence on a CCD-camera. Excess ions are removed by lowering the axial trapping potential with a predefined voltage sequence. The cold ions are extracted along the axial direction of the trap by applying an acceleration voltage ranging from 0 to -6$\,$kV, to one end-cap. This voltage is controlled by a fast solid-state switch with a jitter of less than 1$\,$ns. The extraction time is triggered to the phase of the radio-frequency trap-drive ($\Omega/(2\pi)=23\,$MHz). Extraction rates of single ions of up to 3$\,$Hz, can be achieved, corresponding to an average flux of about 0.5$\,$atto-ampere. The ions leave the trap passing through a hole with a diameter of 200$\,\mu$m in the end-cap. 

Two pairs of deflection electrodes are used for alignment and scanning of the beam. They are located at a distance of 46$\,$mm and 67$\,$mm from the center of the trap (see Figure~\ref{fig:trap}b). In order to focus the ion beam, a electrostatic einzel-lens is placed 332$\,$mm from the trap. It is constructed from three coaxially arranged ring shaped electrodes with an open aperture of 4$\,$mm. In order to minimize spherical aberration, the geometry parameters are optimized with numerical simulations of the electrostatic field~\cite{Singer:2010}. Chromatic aberration is suppressed because of the narrow velocity distribution the of ions. For the time of flight we measure a half width half maximum spread of $\Delta t=270\,$ps. This corresponds to a velocity spread of $\Delta v=8\,$m/s at an typical average speed of about $10^5\,$m/s. Microscopy is implemented by placing a partially transmissive object on a three-axis translation stage. This can be either a nano-structured test sample or a profiling edge. The transmitted ions are detected by a secondary electron multiplier. Image information is generated by recording transmission events for a well defined number of extractions while scanning the object position in the focal plane.

\section{Imaging with single ions}

In electron or ion microscopy poor SNR generally can be overcome by increasing the exposure time or the current. This is a direct consequence of the Poissonian statistics of the sources in use. For some applications however, it is important to minimize the current, as for example where high irradiation causes inconvenient charging~\cite{YoungMin2010Practical}, contamination or even damage~\cite{Prawer1995IonBeam} to delicate samples. However, with conventional Poissonian sources the SNR becomes increasingly worse when going to lower currents. Thus effectively lowering the information gain per particle.

\begin{figure}[ht]
\begin{center}
\includegraphics[height=5cm]{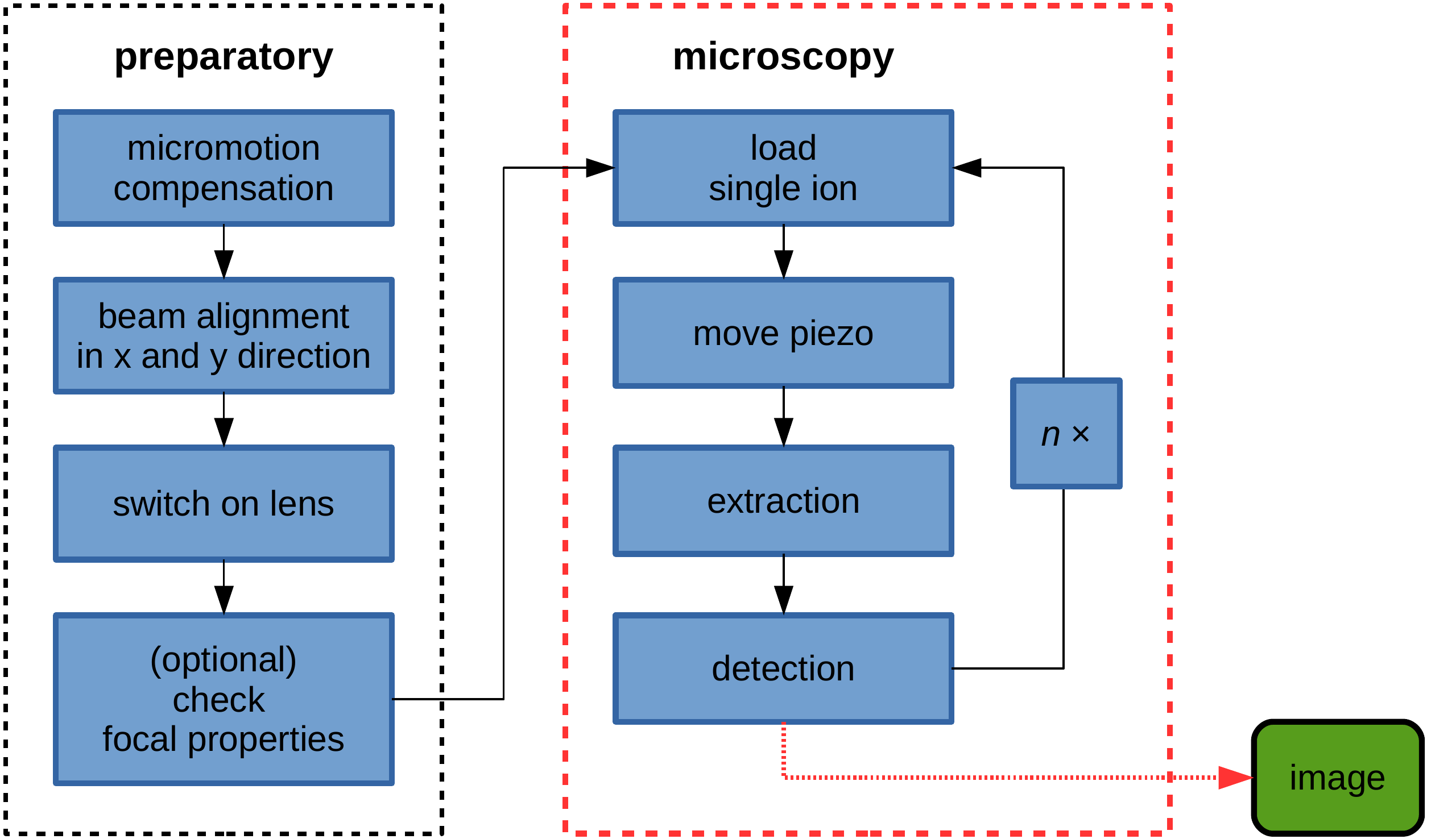}	
\end{center}
\caption{A flowchart of the microscopy experiment, separated into preparatory steps and the microscopy protocol itself.\label{fig:microscopy flowchart}}
\end{figure}

In order to solve this problem we employ a deterministic source, which provides an higher SNR compared to Poissonian sources, especially when approaching the single particle regime. This remains valid even if a detector of finite quantum efficiency is used, leading to Binomial statistics. Figure~\ref{fig:microscopy flowchart} shows how the single ion microscope is implemented in the experiment.

\begin{figure}[ht]
\begin{center}
\includegraphics[height=5cm]{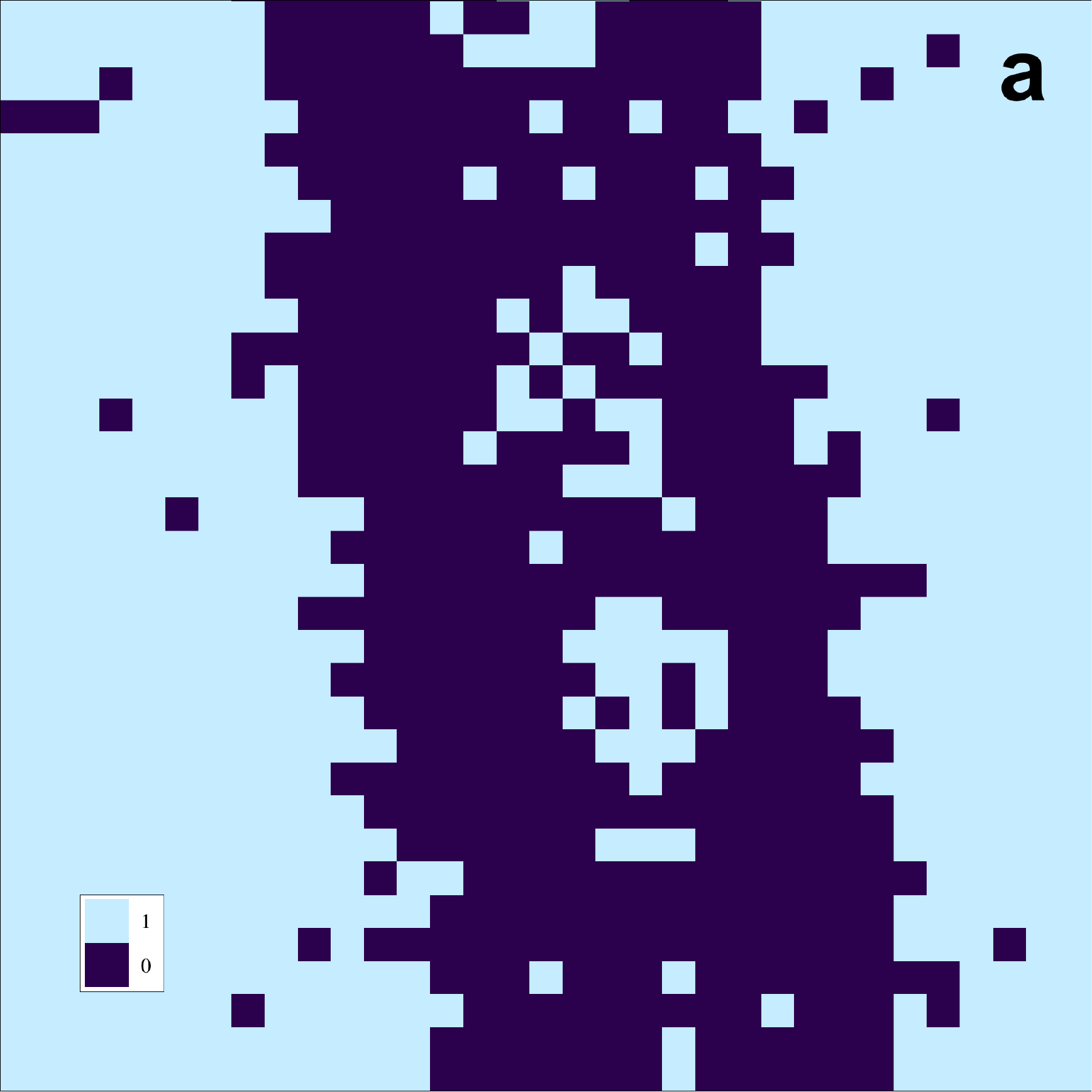}
\includegraphics[height=5cm]{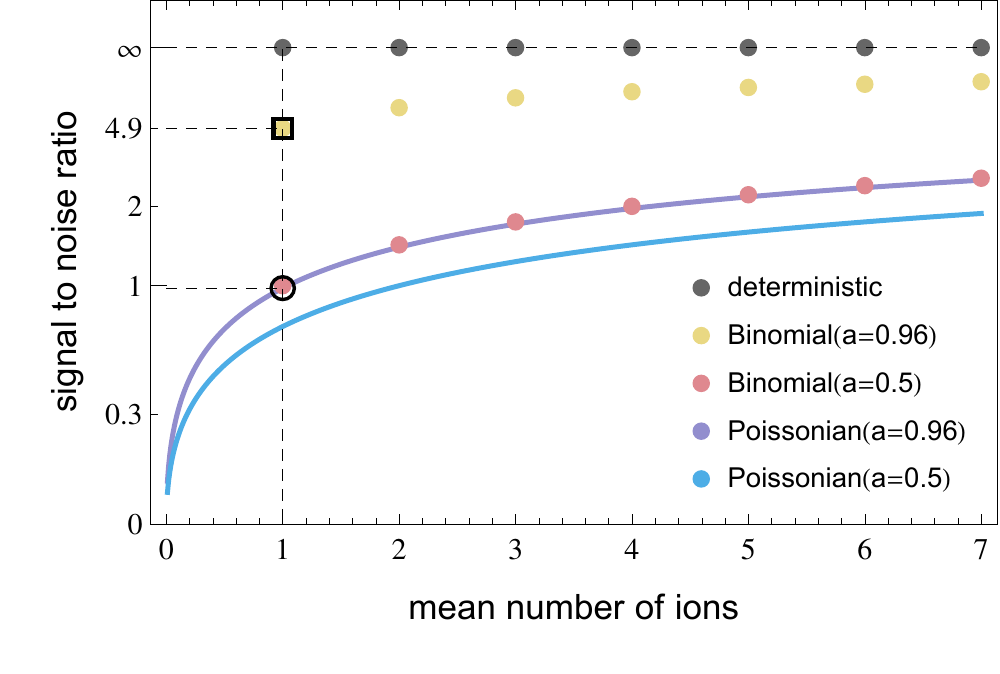}
\end{center}
\caption{a) As an example a waveguide-cavity structure made from diamond is scanned using one ion at each lateral position, with a resolution of (25x25)$\,$nm$^2$ per pixel. The holes have a diameter of about 150$\,$nm.\label{fig:microscopy_example} b) Calculated SNR of a deterministic and a Poissonian source plotted as a function of the mean number of extracted ions for different detector efficiencies. The scale on the y-axis is compactified by the function $f/(f+1)$. Note that the SNR for binomial statistics with $a=0.5$ is identical to the Poissonian SNR with $a=1$. The square depicts the operating point of a deterministic source, with one ion per pixel, and the circle shows the operating point of an equivalent Poissonian source, where the mean number of ions per pixel equals one. The detector efficiency is assumed $0.96$ for both cases. This results in an SNR of $4.90$ for our deterministic source and $0.96$ for the Poissonian, respectively. 
\label{fig:scanplotDiscuss}}
\end{figure}

As an example Figure~\ref{fig:microscopy_example}a) shows the result of an imaging scan of a optical waveguide-cavity structure made from diamond~\cite{riedrich2012one}, using exactly one ion at each position, with a resolution of (25x25)\,nm$^2$.

A quantitative comparison in terms of SNR, of a deterministic sources with a conventional Poissonian source, is presented in Figure~\ref{fig:scanplotDiscuss}b). The SNR, calculated as a function of the mean number of extracted ions, is compared for different detection probabilities $a$. Note that the dark count noise is not taken into account in this comparison. For the plot the definition SNR~$=\mu/\sigma$ is used, where $\mu$ is the mean value and $\sigma$ the standard deviation of the corresponding probability mass function.

\section{Bayes experimental Design}

We use the Bayes experimental design method\cite{lindley1956measure,guerlin2007progressive,pezze2007phase,brakhane2012bayesian} to image transmissive structures with optimal efficiency: The algorithm optimizes the probing position for the next probing event by maximizing the expected information gain. This allows to measure parameter values of one or two-dimensional transmissive structures, which are modelled by parametrizing their contour function. The parameter values are determined by incorporating the measurement results, using the Bayes update rule. The model is given as a convolution of this contour function and the beam profile. First we introduce and explain the Bayesian method on an abstract level. Then an example is given by means of a simple profiling scan, demonstrating how the radius and the position of the beam can be obtained by applying this method. In a second example an algorithm is presented which is able to find and determine the exact lateral position of a circular hole structure with optimal efficiency.

\begin{figure}[ht]
\begin{center}
\includegraphics[height=7cm]{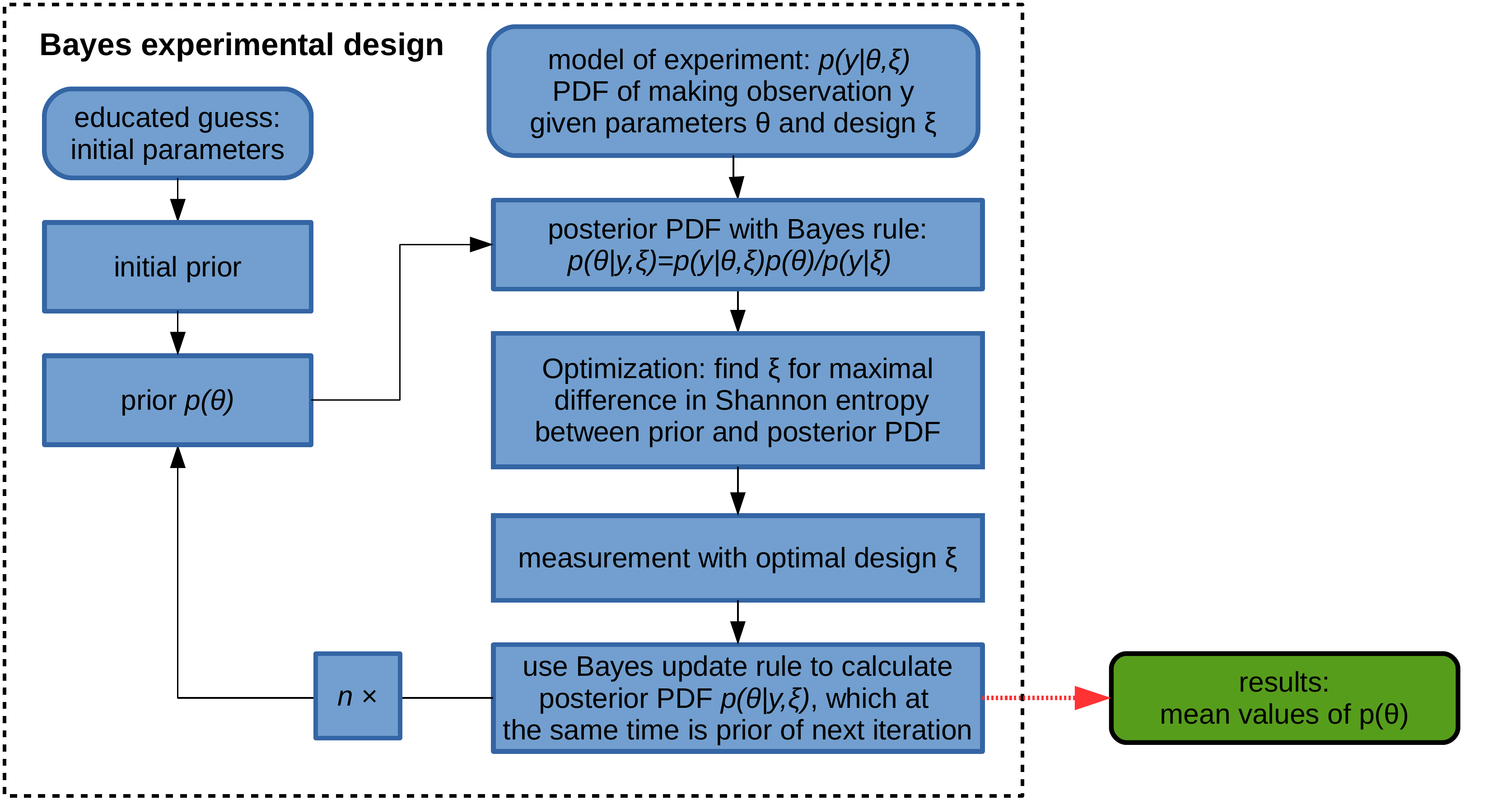}	
\caption{\setlength{\baselineskip}{15pt}{A flowchart explaining the Bayes experimental design method.}}
\label{fig:bayes flowchart}
\end{center}
\end{figure}

In the Bayesian concept of probability, the information or knowledge about the value of a parameter $\theta$, is represented by a probability distribution function (PDF). In the context of measurement, the Bayes update rule allows for subsequently incorporating new information from the outcome $y$ of a measurement, into the \textit{prior} PDF $p(\theta)$, which represents the pre-existing information. The resulting PDF is called the \textit{posterior}:
\begin{equation}
p(\theta|y,\xi)=\frac{p(y|\theta,\xi)p(\theta)}{p(y|\xi)}.
\label{eq:Bayes update rule}
\end{equation}
The right hand side of the equation is the product of the prior PDF and the statistical model of the measurement $p(y\vert\theta,\xi)$, which is the probability to observe an outcome $y$ given the parameter values $\theta$ and design parameters $\xi$. $\xi$ contains the free control parameters of the experiment. Normalization is provided by the marginal probability of observing $y$, 
\begin{equation*}
p(y|\xi)=\int p(y|\theta,\xi)p(\theta) d\theta.
\end{equation*}
The Bayes experimental design method consists in maximizing the information gain per measurement by means of manipulating the design parameters. The information gain of a measurement with outcome $y$ and control parameters $\xi$ is represented by the \textit{utility} $U(y,\xi)$, which is given by the difference between the Shannon entropies of the posterior and the prior PDF:
\begin{equation*}
U(y,\xi)= \int \ln(p(\theta|y,\xi))p(\theta|y,\xi) d\theta-\int \ln(p(\theta))p(\theta) d\theta.
\end{equation*}
We obtain a utility, independent of the hitherto unknown observation, by averaging over the measurement outcomes:
\begin{equation}
U(\xi)=\sum_{y\in\lbrace 0,1 \rbrace} U(y,\xi)p(y|\xi).
\label{eq:utility}
\end{equation}
The $\xi$ value which yields maximal utility is used to carry out the measurement, ensuring optimal information gain.

\begin{figure}[ht]
\begin{center}
\includegraphics[height=5cm]{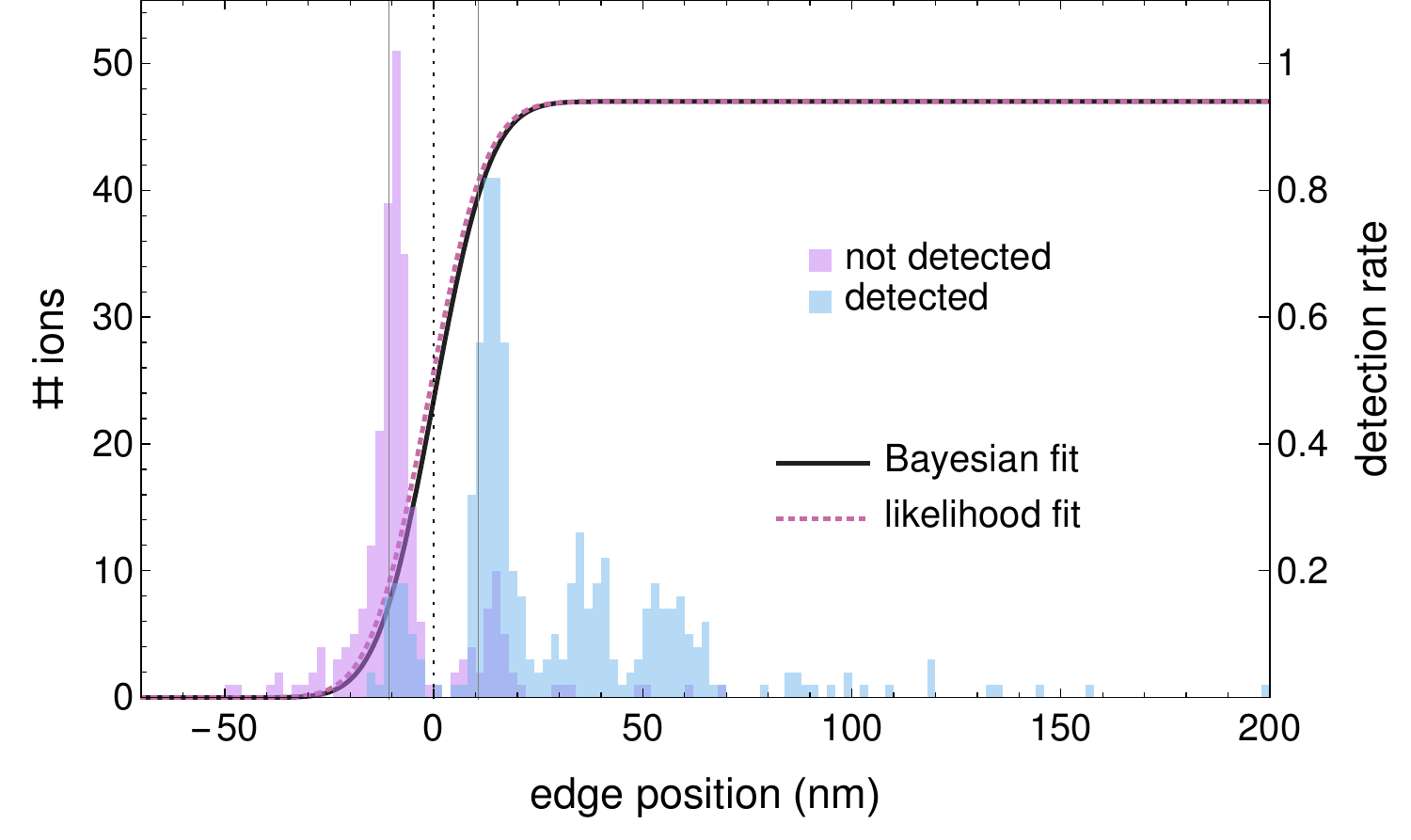}
\end{center}
\caption{Typical beam profiling measurements using the Bayes experimental design method. The histogram resembles the distribution of optimal blade positions as calculated by the algorithm during the measurement. This data is split into cases where the ion was transmitted and cases where it was blocked. The Bayesian fit function $p(1|\theta,\xi)$ is drawn according to the final parameter values ($\sigma=10.74\pm1.31\,$nm, $a=0.95\pm0.02$, $x_0$ is set to zero) which are determined iteratively by the Bayesian update rule. For comparison, the purple function shows the result of a maximum likelihood fit to the entire data after the measurement ($\sigma=11.16\pm1.23\,$nm, $a=0.95\pm0.02$, $x_0=0.73\,$nm), since the values determined by the Bayesian method are not independent of the exact sequence.
\label{fig:focus}
}
\end{figure}

For the example of the profiling edge measurement, the design parameter $\xi$ is the profiling edge position, while the parameters to be determined are the beam position $x_0$, its radius $\sigma$ and the detector efficiency $a$, \textit{i.e.} $\theta=(x_0,\sigma,a)$. The outcome of the measurement is binary, $y=\{0,1\}$. The measurement is modelled as
\begin{equation*}
p(y|\theta,\xi) = \begin{cases} 
\frac{a}{2} \, \text{erfc}\left[ \frac{\xi - x_0}{\sigma \sqrt{2}} \right] & \quad \text{if } y=1\\
1-\frac{a}{2} \, \text{erfc}\left[ \frac{\xi - x_0}{\sigma \sqrt{2}} \right] & \quad \text{if } y=0\\
\end{cases}
\end{equation*}
which in this case is a convolution of the transmission function of the structure to be imaged and a Gaussian beam profile.

The Bayes experimental design routine (see Figure~\ref{fig:bayes flowchart}) is implemented as follows. The initial prior, a three dimensional joint PDF of the parameters, position $x_0$, sigma radius $\sigma$ and  detector efficiency $a$, is calculated from an initial guess of these parameters. The marginals of the prior can be uniform or an educated guess \textit{e.g.} a Gaussian distribution. It is implemented numerically, being a three dimensional grid of equidistant, weighted and normalized sampling points. In order to calculate the utility, the posterior PDF is calculated by applying the Bayes update~(\ref{eq:Bayes update rule}) for each sampling point and the integrals are replaced by sums over all sampling points. The maximizing algorithm is now realized by calculating the utility for equally spaced profiling edge positions within the interval of interest, and recursively repeating this calculation for a smaller interval around the position with the highest utility. Five recursions were found to be sufficient to reach the required accuracy without incurring excessive computational expense. Using the measurement outcome of the real experiment performed at the calculated optimal profiling edge position, the Bayesian update~(\ref{eq:Bayes update rule}) is applied to calculate the actual posterior PDF, which assumes the role of the prior PDF for the next iteration. The procedure is repeated until an accuracy goal is reached.

\begin{figure}[ht]
\begin{center}
\includegraphics[height=5cm]{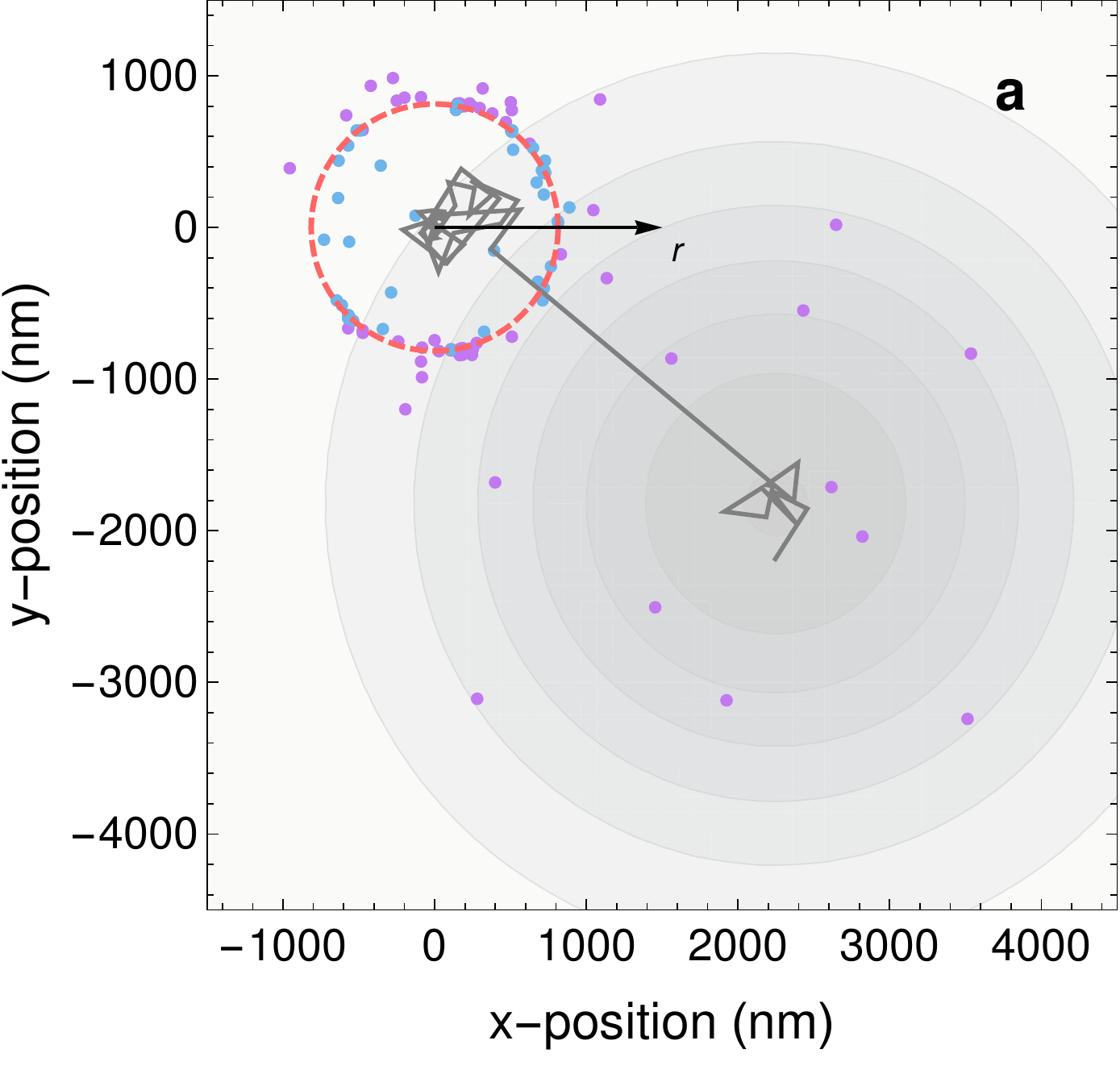}
\includegraphics[height=5cm]{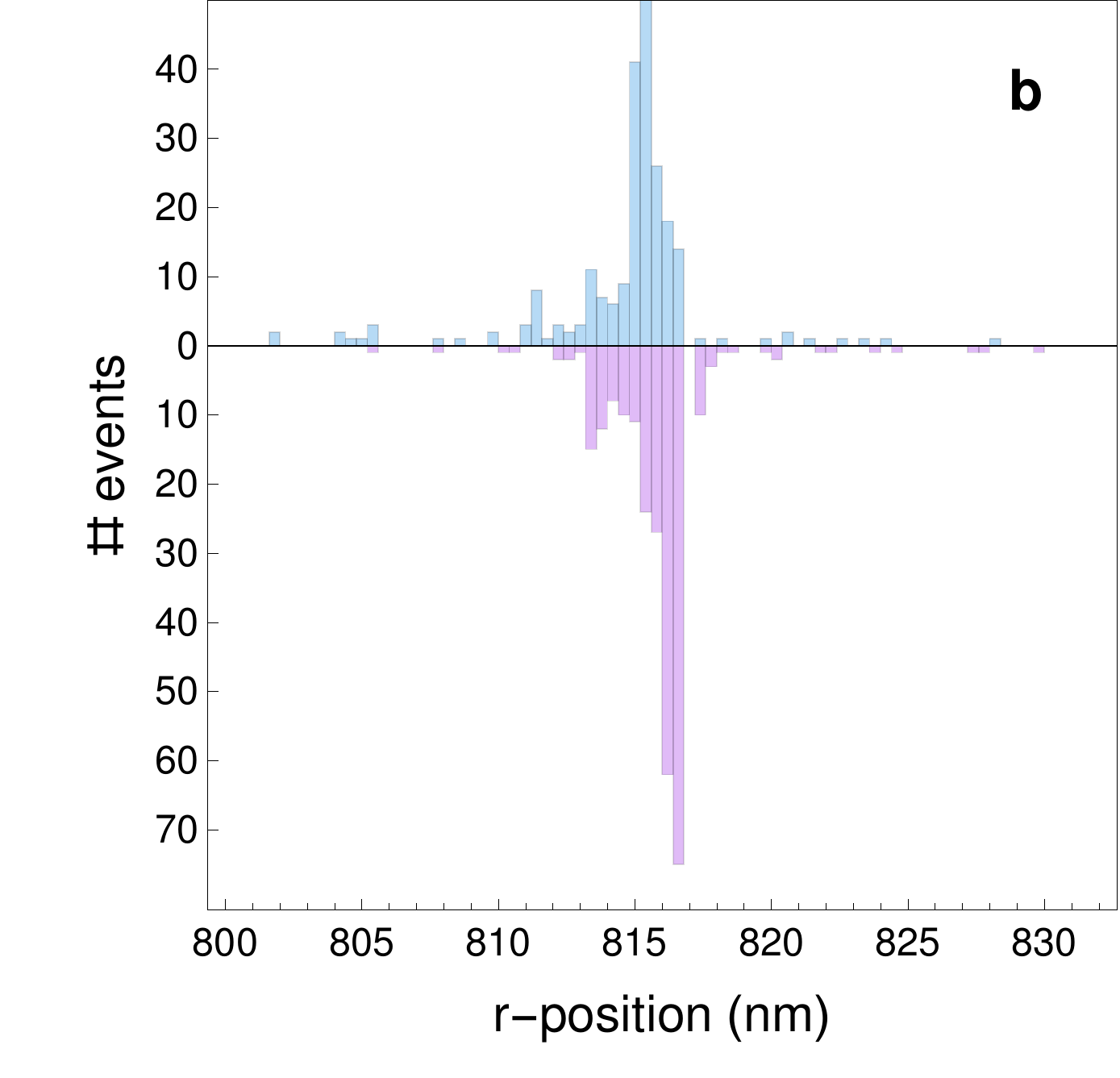}
\caption{a) Determination of alignment-hole parameters with the Bayes method. For the plot the $x$ and $y$ position of the hole were set to zero. The blue and red dots represent the positions where an ion was, or was not, detected, respectively. The final location and radius of the hole structure is depicted by the dashed red circle. The initial guess of the Gaussian shaped PDF for the position is depicted as a gray shade in the background. The dark gray line follows the progression of its mean value \textit{i.e.} the assumed center position of this distribution as a function of the number of extracted ions. Within the first eleven iterations no ion is transmitted. The spatial information of these blocked particles shifts the assumed position, since it excludes that specific areas are transmissive. After the first ion is transmitted the assumed position makes a step towards its location. b) A histogram of detected and not detected events dependent on $r$ the distance to the center of the structure.
\label{fig:BayesMarker}}
\end{center}
\end{figure}
We demonstrate imaging of two-dimensional transmissive structures with a parametrizable transmission function, by determining the parameter values of a circular hole in a diamond sample (see Figure~\ref{fig:BayesMarker}) using the Bayes experimental design method. This is also a practical example for sample alignment, since for many applications it is useful to know the exact lateral position of a sample with respect to the beam focus. For this purpose two perpendicular profiling edges as used in the beam radius measurement could equally be employed. However, for practical reasons, it might be more convenient to use a simple hole structure as a marker, which is in close proximity to a structure of interest.

The experiment is parametrized by the lateral position of the center of the circular hole, its radius as well as the 1$\,\sigma$-radius of the ion beam and the detector efficiency. The radius of the beam and the detector efficiency were kept constant at 25$\,$nm and 95$\,$\% respectively. Both values were measured separately in advance. Using 572 ions in total, the position was determined with an accuracy of $\Delta x=3.5\,$nm and $\Delta y=2.0\,$nm, where the radius was measured to be $r=814.1\pm1.5\,$nm. The systematic errors resulting from the deviations of the shape to the parametrization (ideal circle) are difficult to quantify, since the precise extent of this deviation is unknown. However, the accuracy of the results apply to an ideal circular shape, which could be available in other experiments.

\section{outlook: single ion implantation}

\begin{figure}[ht]
\begin{center}
\includegraphics[height=7cm]{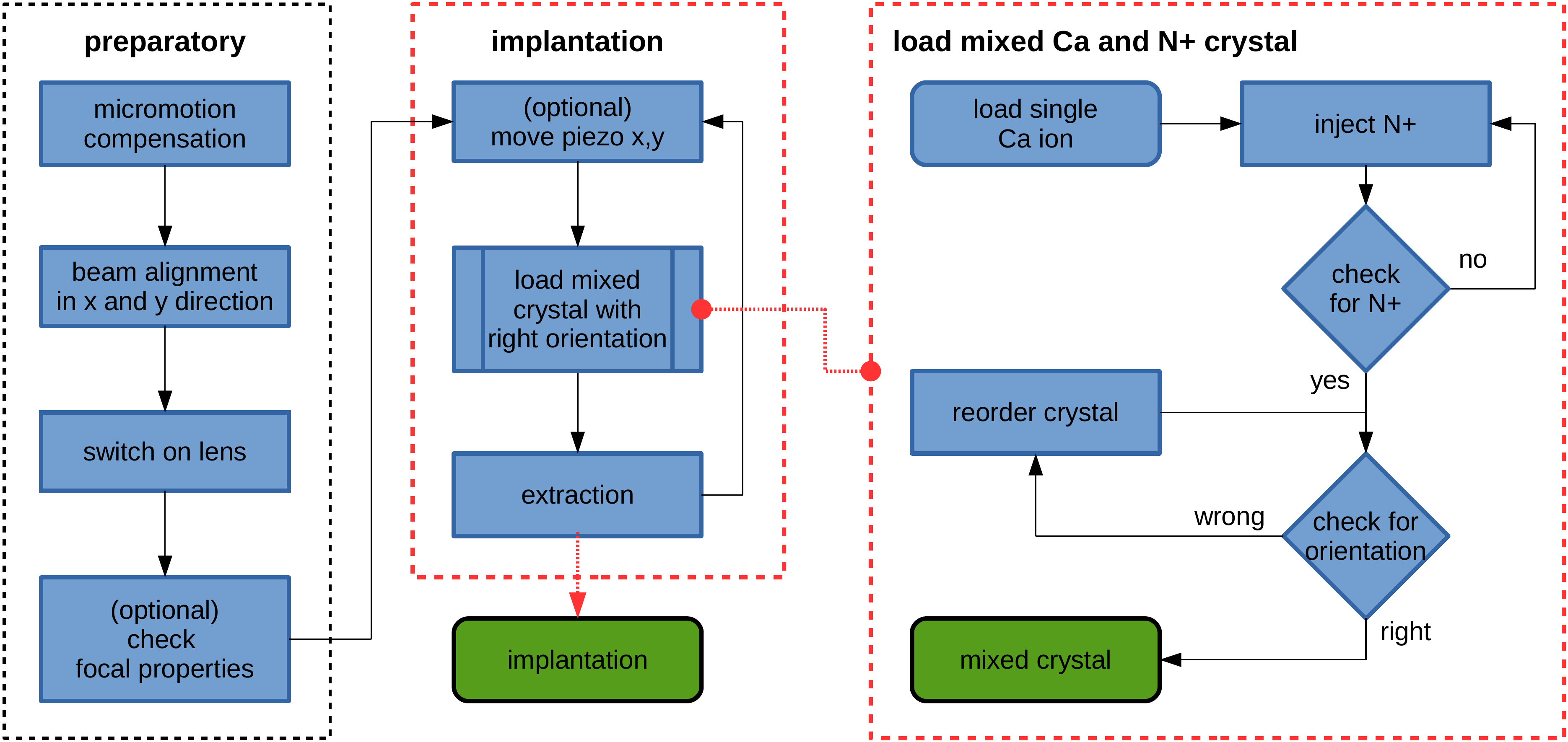}	
\caption{\setlength{\baselineskip}{15pt}{A flowchart of the implantation protocol, separated into preparatory steps and the implantation protocol. A detailed chart of the loading of mixed ion crystals is shown seperately on the right hand side.}}
\label{fig:implantation flowchart}
\end{center}
\end{figure}

Besides imaging, the apparatus is also designed for deterministic ion implantation (see Figure~\ref{fig:implantation flowchart}) on the nanometer scale. This would allow for the fabrication of scalable solid state quantum devices, where individual dopants are coupled by their mutual dipolar magnetic interaction. Our focus is on systems of coupled nitrogen vacancy color centres~\cite{dolde2013room}, coupled single phosphorous nuclear spins in silicon~\cite{kane1998silicon,jamieson2005controlled,pla2013high,veldhorst2015twoQubit} and cerium or praseodymium in yttrium orthosilicate~\cite{kolesov2012optical}. Another promising application of a highly focussed deterministic single ion beam is the doping and structuring of graphene~\cite{kotakoski2015toward}. 

For applications of ion implantation where absolute positioning is necessary, transmissive structures can be used for referencing. Imaging these markers with the same source which is used for the implantation, allows for the accurate alignment of dopants relative to this markers, free of parallax errors. Depending on the application, it can be necessary to use a different ion species for the imaging to avoid contamination of the sample. For this purpose, it has to be considered that the energy of the different species remain the same, in order to avoid different spatial positions of the beam focus as a result of inconsistent deflection.

\acknowledgments 
 
The authors acknowledge discussions with S. Prawer and G. Sch\"onhense. The project acknowledges financial support by the Volkswagen-Stiftung, the DFG-Forschergruppe (FOR 1493) and the EU-projects DIAMANT and SIQS (both FP7-ICT). FSK thanks for financial support from the DFG in the DIP program (FO 703/2-1).

\end{document}